\documentclass{article}

\usepackage{arxiv}

\usepackage[utf8]{inputenc} 
\usepackage[T1]{fontenc}    
\usepackage{hyperref}       
\usepackage{url}            
\usepackage{booktabs}       
\usepackage{amsfonts}       
\usepackage{nicefrac}       
\usepackage{microtype}      
\usepackage{lipsum}		
\usepackage{graphicx}
\usepackage{natbib}
\usepackage{doi}
\usepackage{amsmath}
\usepackage{amssymb}

\title{Dispersion--Guided Physics--Aware Deep Inverse Operator for Surface Wave Mode Separation}

\author{
\href{https://orcid.org/0009-0008-4685-9458}
{\includegraphics[height=1.6ex]{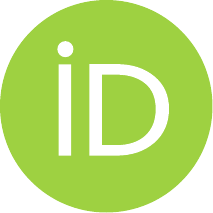}\hspace{1mm}Yang Cui}%
\thanks{Corresponding author:
\href{mailto:yang.cui@geo.uu.se}{\texttt{yang.cui@geo.uu.se}}}\\
Department of Earth Sciences\\
Uppsala University\\
752 36 Uppsala, Sweden
\And
Sujith Swaminadhan\\
Bureau of Economic Geology\\
Jackson School of Geosciences\\
The University of Texas at Austin\\
Austin, Texas 78713-8924, USA
\And
Yangkang Chen\\
Bureau of Economic Geology\\
Jackson School of Geosciences\\
The University of Texas at Austin\\
Austin, Texas 78713-8924, USA
\And
Christian Schiffer\\
Department of Earth Sciences\\
Uppsala University\\
752 36 Uppsala, Sweden
\And
Myrto Papadopoulou\\
Department of Earth Sciences\\
Uppsala University\\
752 36 Uppsala, Sweden
}

\renewcommand{\shorttitle}{Deep Inverse Operator}

\hypersetup{
pdftitle={A template for the arxiv style},
pdfsubject={q-bio.NC, q-bio.QM},
pdfauthor={David S.~Hippocampus, Elias D.~Striatum},
pdfkeywords={First keyword, Second keyword, More},
}

\begin{document}
\maketitle

\begin{abstract}
	Surface-wave (SW) dispersion analysis is widely used in near-surface geophysics and seismology to determine shear-wave velocity structures by measuring SW geometric dispersion in seismic data. Among the available approaches, multi-channel analysis of surface waves (MASW) and two-station methods are commonly employed to extract the dispersion information for SW inversion. However, the coexistence of fundamental and higher modes in seismic data poses challenges for the method, particularly in two-station analysis. To separate the different mode components, we propose a physics-aware unsupervised deep learning framework. The method acts as a deep inverse operator that directly separates fundamental- and higher-mode components in the time–space domain using an adaptive Gaussian mask constructed in the frequency–phase-velocity ($f$--$v$) domain. Physical constraints are incorporated into the loss function by maximizing energy concentration within the target mask while suppressing leakage outside it. Through backpropagation, the network learns the inverse mapping from the $f$--$v$ domain physical constraints to time–space domain wavefield separation without requiring labeled training data. Numerical experiments on both synthetic and field data show that the framework provides a robust and automated solution for SW mode separation, facilitating more reliable dispersion curve picking and improving the accuracy of subsequent SW inversion.
\end{abstract}

\keywords{Surface-Wave \and Deep Learning \and Inverse Problem \and Signal Processing \and Seismic}

\section{Introduction}
Surface-wave (SW) methods are widely used to investigate the shear-wave velocity structure of the subsurface across scales ranging from near-surface characterization (e.g., geotechnical investigations, environmental studies) and exploration~\citep{xia1999estimation,papadopoulou2023high} to regional structure imaging~\citep{dorman1962numerical,socco2010surface}. 

A key step in these methods is the accurate measurement of dispersion curves (DCs). Among the available techniques, multi-channel analysis of surface waves (MASW) is one of the most common methods for processing active-source surface-wave data, which transforms multichannel shot gathers into dispersion images and extracts DCs for subsequent inversion~\citep{park1998imaging,park1999multichannel}. Alternatively, two-station methods, originally introduced in seismology, provide a means of estimating inter-station phase velocities. Although they have been widely used in SW tomography, particularly when dispersion measurements are required along multiple receiver pairs~\citep{yao2006surface,yao2008surface,gosselin2023probabilistic}, the same principle can also be applied to active-source data~\citep{ikeda2020two}. The main difference between these workflows lies in how dispersion information is obtained: MASW estimates dispersion from the wavefield recorded by an array, whereas two-station methods estimate phase velocity from the phase difference between receiver pairs. In both cases, the quality of the dispersion measurements directly affects the reliability of the inverted shear-wave velocity models~\citep{papadopoulou2021surface}.

Regardless of the processing method used, accurate picking of physically meaningful DCs is essential for robust inversion and reliable subsurface characterization. In many SW applications, the fundamental mode dominates the recorded wavefield and appears as a clear energy maximum in $f$--$v$ dispersion images, making it relatively straightforward to identify and pick. For this reason, fundamental-mode analysis forms the basis of many common SW methods~\citep{foti2014surface}. However, field data may also contain higher modes, body waves, and noise. Although these components may appear as distinct energy trends in spectral images, their interference and pronounced amplitude differences can obscure modal identification, make DC picking ambiguous, and complicate automated picking procedures~\citep{chamorro2024deep}. This ambiguity is particularly problematic for two-station SW analysis because the limited spatial sampling makes it difficult to separate mixed modal distribution. As a result, mixed-mode contamination can lead to unstable and biased phase-velocity measurements.

Although the fundamental mode is commonly used in routine inversion, higher modes can also provide valuable constraints. According to~\cite{xia2003inversion}, higher-mode components increase depth sensitivity and improve the resolution of the inverted shear-wave velocity model, especially for structures with strong velocity contrasts or low-velocity layers~\citep{pan2019sensitivity}. Multimodal inversion of SW can therefore be beneficial when the different-mode components are correctly identified. Separating modal components before dispersion analysis can improve both fundamental-mode workflows and multimodal inversion. Several transform-domain methods have been proposed to enhance dispersion imaging and to separate modal energy. These methods usually exploit the fact that different SW modes occupy different regions in the frequency-wavenumber or $f$--$v$ domain. For example, high-resolution linear Radon transform has been used for Rayleigh-wave dispersion imaging and mode separation~\citep{luo2008rayleigh}. However, with field data, the inverse Radon transform will suffer from artifacts due to its sensitivity to spatial sampling and acquisition conditions~\citep{schonewille2001parabolic}. In addition, sharp filtering followed by inverse transformation may introduce artifacts in the reconstructed space-time wavefield. \cite{jin2025extracting} utilized the high-resolution linear Radon transform to extract multimodal DCs from ambient noise data, where they showed that their method can reduce the inversion non-uniqueness by leveraging both fundamental-mode and higher-mode curves~\citep{cercato2009addressing}. These limitations motivate a more flexible mode-separation framework that preserves dispersion-domain physical constraints while reducing reconstruction artifacts.

Deep learning (DL) learns high-order, non-linear features between the training domain and the target domain using large volumes of paired data~\citep{geofwi}. Over the past decade, it has been widely used in seismology for tasks such as denoising seismic data, reconstructing missing traces, picking earthquake phases, and full-waveform inversion~\citep{omar2020ddae,shucai2020,mousavi2022deep,chao20252,yang2026unsupervised}. For SW exploration, DL offers an alternative way to represent the complex mappings between mixed and separated wavefields. \cite{li2020separation} proposed a supervised DL method for separating multi-mode SWs, demonstrating that neural networks can learn mode-dependent features from a large volume of training samples. For automatic DC picking, \cite{chamorro2024deep} proposed a DL method to learn the mapping from shot gathers to fundamental-mode and first-mode curves. Moreover, they explore ways to achieve quality control. However, the performance of the supervised method often relies on high-quality labeled data, which limits its robustness where it is hard to generate labels for field data. Such labels are difficult to obtain from field data and may be imperfect even in synthetic examples. Moreover, a purely data-driven model may learn statistical correlations in the training set without explicitly enforcing the dispersion physics governing SW propagation.

Physics-aware DL offers an effective way to reduce the dependence on labeled data. Instead of training a network only with paired input-output samples, physical knowledge can be embedded in the loss function, thereby acting as a constraint on the optimization. Inspired by~\cite{cui2025learning}, where the authors used physics-aware loss in $f$--$k$ domain as a correction term in supervised training, here we used the physics-aware loss in $f$--$v$ domain purely driven by an unsupervised loss function. For SW mode separation, the dispersion images provide a natural physical prior. Each mode is associated with a characteristic trend in the $f$--$v$ domain, guiding the separation of modal components in the original space-time domain. Compared with conventional dispersion-domain masking, the proposed network-based reconstruction provides a more flexible inverse mapping while reducing artifacts caused by sparse sampling and sharp filtering. Compared with supervised DL, the proposed framework does not require clean modal labels, making it better suited to field data applications. The method is evaluated using both synthetic and field seismic data from an Alum Shale exploration project in Sweden. The results show that the separated fundamental-mode component yields cleaner dispersion images and more stable DC picking, benefiting MASW, two-station phase-velocity measurements, and subsequent SW inversion and tomography.

\section{Theory}
This section describes the theoretical principles of the proposed dispersion-guided, physics-aware DL framework. We first introduce the theory of SW dispersion analysis. We then formulate the generation of the adaptive Gaussian mask in the $f$--$v$ domain. Finally, we present the DL architecture and the associated physics-aware loss functions.

\subsection{$f$--$v$ analysis of SW data}
SW $f$--$v$ analysis is widely used to estimate Rayleigh-wave DCs using MASW. The method converts the input seismic data from space-time domain to $f$--$v$ domain, where SWs correspond to high-energy ridges, possibly in multiple modes. 
\begin{figure}
    \centering
    \includegraphics[width=0.9\linewidth]{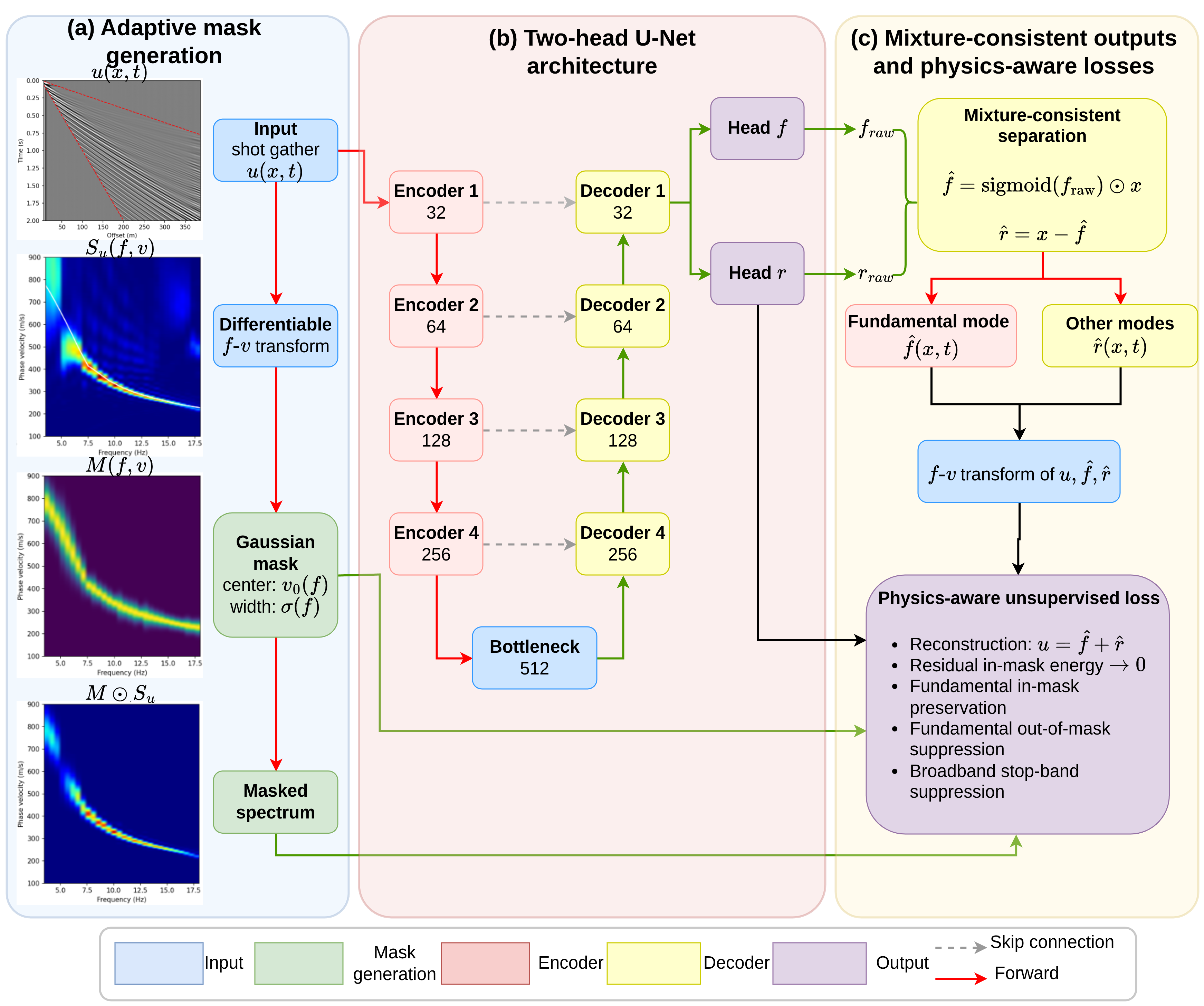}
    \caption{Diagram of mode separation workflow. (a) Workflow of adaptive mask generation: it begins with the automatic DC maxima picking, followed by the Gaussian mask generation based on the picked DC. Finally, we multiply the mask with the $f$--$v$ dispersion image to get the energy of the fundamental mode. (b) Two-head U-Net architecture for mode separation, where the first output is the fundamental mode and the second one is the higher-mode in the space-time domain. (c) The physics-aware loss function, which embeds the dispersion information into the loss function.}
    \label{fig:mode_sep_workflow}
\end{figure}

Given the angular frequency $\omega=2\pi f$ and the phase velocity $v=\omega/k$. A monochromatic SW can be described by: 
\begin{equation}
    u(x,t) = Ae^{i(\omega t-kx)},
    \label{eq:fv1}
\end{equation}
where $A$ is the amplitude, $k$ is the wavenumber, and $f$ represents the frequency. 

The geometric dispersion of Rayleigh waves means that different frequency components travel at different velocities. To measure this, the signal is transformed from the time domain to the frequency domain using the Fourier transform: 
\begin{equation}
   U(x, f) = \int_{-\infty}^{\infty}{u(x,t)e^{-i2\pi ft}dt}.
\label{eq:fv2}
\end{equation} 
For a propagating wave, each frequency component contains a spatial phase term. Assuming propagation in the positive $x$ direction and neglecting attenuation, the frequency-domain wavefield can be written as:
\begin{equation}
    U(x,f) = \tilde{A}(f)e^{-ik(f)x}.
\label{eq:fv3}
\end{equation}

\subsection{Adaptive masking strategy}
Figure~\ref{fig:mode_sep_workflow}a illustrates the construction of the adaptive Gaussian mask. Starting from the input shot gathers, we apply $f$-$v$ analysis~\citep{park1999multichannel} to obtain the dispersion image. The fundamental-mode DC is automatically picked by searching for the energy maxima within a selected frequency band. A Gaussian function centered on the picked DC is generated, with a predefined bandwidth that defines the modal acceptance region.

Let $u\in\mathbb{R}^{T\times N_r}$ denote the input shot gather, and let $\hat f,\hat r\in\mathbb{R}^{T\times N_r}$ be the network outputs (fundamental and residual modes, respectively). On the input shot gather $u$, a forward $f$--$v$ transform $\mathcal{T}(\cdot)$ is applied,
\begin{equation}
S_u(v,f)
=
\left|
\mathcal{T}\!\left(u_{:\,,\Omega}\right)(v,f)
\right|
\in\mathbb{R}_{\geq0}^{N_v\times N_f},
\qquad
S_{\hat f}(v,f)
=
\left|
\mathcal{T}\!\left(\hat f_{:\,,\Omega}\right)(v,f)
\right|
\in\mathbb{R}_{\geq0}^{N_v\times N_f}.
\label{eq:mask1}
\end{equation}
where $\Omega$ denotes the selected receiver-offset range used in the $f$--$v$ transform (only for synthetic data).

From $S_u$, the fundamental DC $v_{\mathrm{dc}}(f)$ is automatically picked by searching for maxima along the velocity axis, and a frequency-dependent bandwidth $\sigma(f)$ is selected based on the local energy spread. The Gaussian mask is then constructed as
\begin{equation}
M(v,f)=
\begin{cases}
\displaystyle
\exp\!\left[
-\frac{\left(v-v_{\mathrm{dc}}(f)\right)^2}
{2\sigma(f)^2}
\right],
&
f\in
\left[
f_{\min}^{\mathrm{pick}},
f_{\max}^{\mathrm{pick}}
\right],
\\[8pt]
0,
&
\text{otherwise}.
\end{cases}
\label{eq:mask2}
\end{equation}
and we define the complementary mask $M_{\mathrm{out}}(v,f)=1-M(v,f)$.

\subsection{Deep learning principle}
U-Net~\citep{ronneberger2015u} is a representative encoder-decoder network that was originally developed for image segmentation and has since been widely used for dense prediction tasks. As shown in Figure~\ref{fig:mode_sep_workflow}b, the proposed framework adopts a two-head U-Net architecture to separate SW modal components in shot gathers. The input to the network is the shot gather in the space-time domain. The encoder extracts multiscale features through a sequence of convolutional blocks, each consisting of two convolutional layers followed by batch normalization and a ReLU activation function. Max-pooling is used after each encoder block to progressively reduce the spatial resolution and enlarge the receptive field, while the number of feature channels is increased to enhance the representation capacity.

At the lowest resolution, a bottleneck block is introduced to capture high-level semantic features and global wavefield patterns. The decoder then restores the spatial resolution through transposed convolutions. At each decoding level, skip connections are used to concatenate the upsampled features with the corresponding encoder features. These skip connections help preserve local waveform details. The final shared decoder features are passed to two separate output heads, which simultaneously predict the fundamental- and higher-mode components. In this way, the network learns a common representation of the mixed wavefield while allowing each head to focus on a specific modal component. This design is well suited for mode separation because the fundamental and higher modes are physically related but exhibit different behavior both in the space-time and $f$--$v$ domains.
  
The proposed method is trained in an unsupervised manner by combining time-domain reconstruction consistency with physics-aware constraints in the $f$--$v$ domain. Given an input shot gather $u(x,t)$, the two-head U-Net predicts fundamental-mode component $\hat{f}(x,t)$ and the higher-mode component $\hat{r}(x,t)$. To ensure mixture consistency, the output is constrained by:
\begin{equation}
    \hat f(x,t) = \sigma\left(f_{\mathrm{raw}}(x,t)\right) \odot u(x,t),
\qquad
\hat r(x,t) = u(x,t) - \hat f(x,t),
\end{equation}
where $\sigma(\cdot)$ denotes the sigmoid function and $\odot$ represents element-wise multiplication. This formulation guarantees that the separated components are consistent with the input mixture.

As shown in Figure~\ref{fig:mode_sep_workflow}(c), the physics-aware constraints are imposed in the $f$--$v$ domain. From Equations~\ref{eq:fv1}--\ref{eq:fv3}, the input gather and the predicted components are transformed using a differentiable phase-shift $f$--$v$ transform~\citep{park1998imaging}, resulting in the dispersion spectra $S_u(f,v)$, $S_f(f,v)$, and $S_r(f,v)$ for the input, predicted fundamental mode, and residual component, respectively. The adaptive Gaussian mask $M(f,v)$ is constructed according to Equations~\ref{eq:mask1} and~\ref{eq:mask2}, which aim to cover the expected fundamental-mode energy region.


The time-domain reconstruction loss is defined as: 
\begin{equation}
    \mathcal{L}_{\mathrm{rec}}=\frac{1}{|\mathcal{I}|}\sum_{(x,t)\in\mathcal{I}}\left|
    u(x,t)-\left(\hat f(x,t)+\hat r(x,t)\right)
    \right|,
\end{equation}
where $\mathcal{I}$ denotes all samples in the input data, and $|\mathcal{I}|=N_tN_r$ is the total number of samples.

To prevent the higher-mode component from containing fundamental-mode energy, the residual energy inside the adaptive mask is minimized by:
\begin{equation}
    \mathcal{L}_{\mathrm{in0}}=\frac{\displaystyle \sum_{f,v} M(f,v) S_r(f,v)}{\displaystyle \sum_{f,v} S_r(f,v) + \varepsilon},
\end{equation}
where $\varepsilon$ is a constant added for numerical stability. This loss measures the fraction of residual spectral energy located inside the fundamental-mode mask.

The predicted fundamental-mode component is encouraged to preserve that input dispersion distribution inside the mask. Following the implementation, the dispersion spectra are first normalized along the velocity axis. This normalization converts the spectrum at each frequency into a relative velocity distribution, thereby reducing the influence of frequency-dependent amplitude variations. 
\begin{equation}
    P_u(f,v)=\frac{S_u(f,v)}{\displaystyle \sum_v S_u(f,v) + \varepsilon},
    \qquad
    P_f(f,v) = \frac{S_f(f,v)}{\displaystyle \sum_v S_f(f,v) + \varepsilon}.
\end{equation}

Let $N_f$ and $N_v$ denote the numbers of discrete frequency and phase velocity samples in the $f$--$v$ dispersion image, respectively. The in-mask preservation loss is then given by
\begin{equation}
    \mathcal{L}_{\mathrm{fkeep}}=\frac{1}{N_f N_v}\sum_{f,v}\left|M(f,v)\left[P_f(f,v)-P_u(f,v)\right]\right|.
\end{equation}

To suppress the leakage from higher modes and noise into the predicted fundamental-mode component, the energy outside the reliable mask region is penalized. A stop-band weight function is defined as: 
\begin{equation}
    M_{\mathrm{stop}}(f,v)=\mathrm{clip}\left(\frac{\tau - M(f,v)}{\tau}, 0, 1 \right)
\end{equation}
where $\tau$ is masked threshold (a constant value). The out-of-mask fundamental suppression loss is: 
\begin{equation}
    \mathcal{L}_{\mathrm{fout}}=\frac{\displaystyle \sum_{f,v} M_{\mathrm{stop}}(f,v) S_f(f,v)}{\displaystyle \sum_{f,v} S_u(f,v)+\varepsilon}.
\end{equation}

In addition, the picked Gaussian mask is projected onto a broader $f$--$v$ grid to form a projected stop mask $M_{\mathrm{stop}}^{\mathrm{proj}}(f,v)$. This term suppresses undesired fundamental-mode energy outside the reliable mask band, especially at frequencies not covered by the picked DC:
\begin{equation}
\mathcal{L}_{\mathrm{fstop}}=\frac{\displaystyle \sum_{f,v}
M_{\mathrm{stop}}^{\mathrm{proj}}(f,v) S_f^{\mathrm{broad}}(f,v)}{\displaystyle \sum_{f,v} S_u^{\mathrm{broad}}(f,v) + \varepsilon},
\end{equation}
where $S_f^{\mathrm{broad}}$ and $S_u^{\mathrm{broad}}$ are computed on the broader $f$--$v$ grid.

The final training objective is therefore:
\begin{equation}
\mathcal{L}
=\lambda_{\mathrm{rec}} \mathcal{L}_{\mathrm{rec}}+\lambda_{\mathrm{in0}} \mathcal{L}_{\mathrm{in0}}+\lambda_{\mathrm{fkeep}} \mathcal{L}_{\mathrm{fkeep}}+\lambda_{\mathrm{fout}} \mathcal{L}_{\mathrm{fout}}+\lambda_{\mathrm{fstop}} \mathcal{L}_{\mathrm{fstop}}.
\end{equation}

In the implementation, these terms correspond to the reconstruction loss, residual in-mask suppression loss, fundamental in-mask preservation loss, fundamental out-of-mask suppression loss, and broadband stop-band suppression loss, respectively. This design allows the network to learn mode separation directly from the input shot gather without requiring labeled fundamental-mode and higher-mode training pairs. In the following synthetic examples, the weights of each loss term are set to 1.0, 1.0, 0.5, 0.5, and 0.2, respectively, selected based on their performance. 

\section{Results}
Based on the two-head U-Net architecture, we formulate an unsupervised, physics-aware loss function to separate the fundamental mode from higher modes in the shot-gather domain. The two output heads of the network correspond to the predicted fundamental-mode component and the residual component containing the remaining modes. This section presents results from both synthetic and field data.

\subsection{Tunable parameters}
The network-related hyperparameters are kept fixed as listed in Table~\ref{tab:hyperparameters}. Additional hyperparameters such as minimum and maximum allowed Gaussian mask width, smoothing window applied to the picked DC before mask generation, and the velocity window around the picked DC used to estimate the adaptive Gaussian width from the input dispersion energy are adjusted based on the input $f$--$v$ dispersion image to balance fundamental-mode coverage and leakage from higher modes. All experiments are performed on an NVIDIA RTX 4090 GPU. For each input gather, the two-head U-Net is optimized for 2,000 epochs using the Adam optimizer, with a typical runtime of approximately 30 s for both synthetic and field examples. This computational efficiency suggests that the proposed method is practical for large-scale seismic-data processing.

\begin{table}[htbp]
\centering
\caption{Hyperparameters setup}
\label{tab:hyperparameters}
\begin{tabular}{ll}
\toprule
\textbf{Parameter} & \textbf{Value} \\
\midrule
Encoder channels & 32, 64, 128, 256 \\
Decoder channels & 256, 128, 64, 32 \\
Activation function & ReLU \\
Optimizer                  & Adam                  \\
Learning rate               & $2\times10^{-4}$      \\
Epochs            & 2000                   \\
\bottomrule
\end{tabular}
\end{table}

For the first step of the workflow (Figure 1a), the maxima of the $f$--$v$ dispersion image were automatically picked within a selected frequency band to obtain stable fundamental-mode picks. For field data, automatic picking proved less reliable due to noise and mode interference, so we manually picked the DCs and used them as the centerline for adaptive Gaussian mask generation. The full frequency range of the manual DCs is used, with minimum and maximum cut-off frequencies set to the first and last picked frequencies, respectively. 

\subsection{Synthetic test}
\begin{figure}
    \centering
    \includegraphics[width=0.6\linewidth]{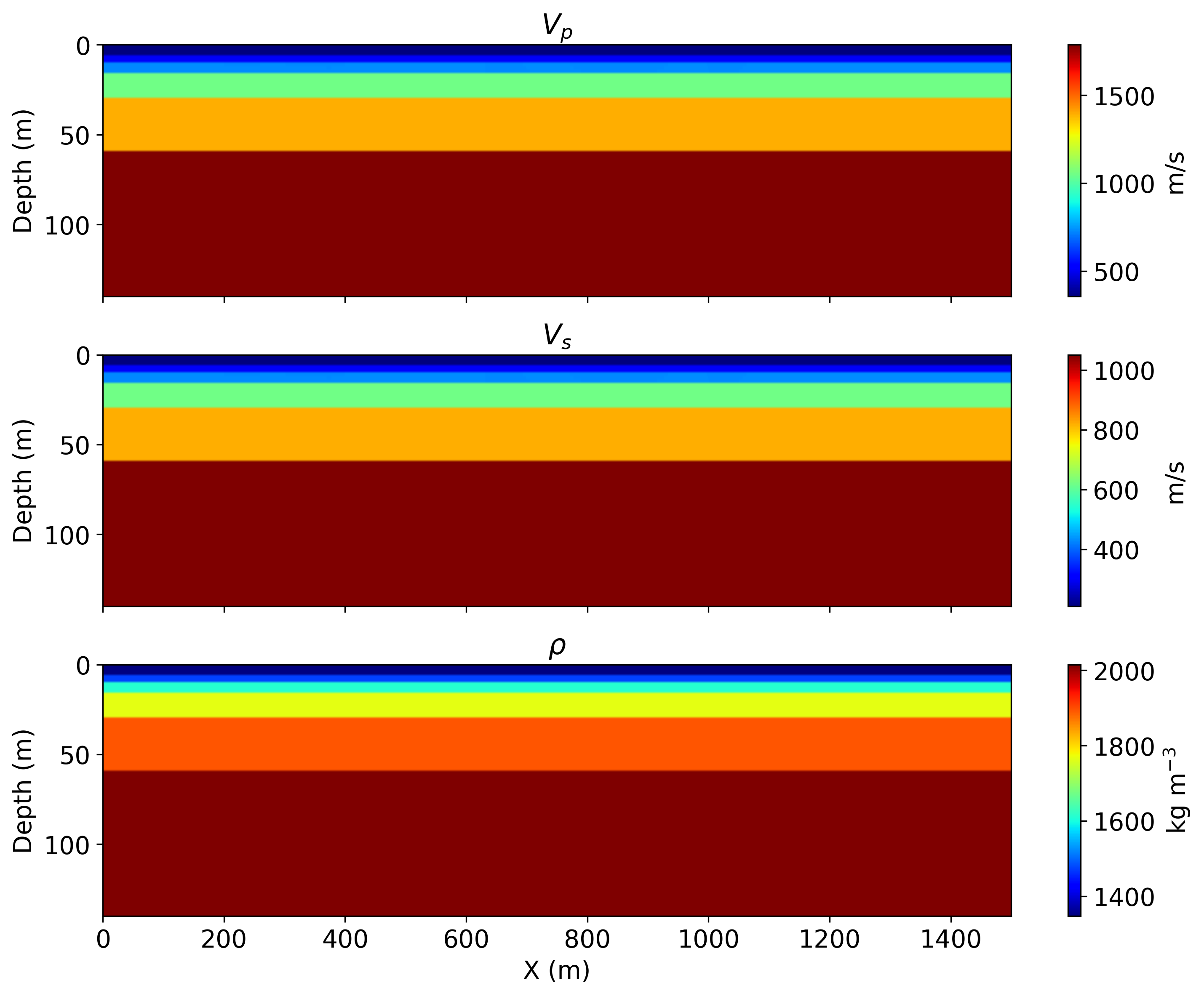}
    \caption{Synthetic layered model for numerical simulation. The $V_p$ model was fixed, while the $V_s$ and $\rho$ models were calculated using empirical equations. This layered configuration allows the numerically obtained higher-mode energy.}
    \label{fig:syn_vel_model}
\end{figure}

To quantitatively assess the performance of the proposed network, we conduct a synthetic experiment based on a layered model. The layered model is selected because it enables a direct comparison between the $f$--$v$ dispersion image obtained from numerical simulation and the theoretical DCs. For synthetic modeling, we used isotropic finite-difference elastic wave equations to simulate the wavefields. The properties of the layered model, including $V_p$, $V_s$, density ($\rho$) are shown in Figure~\ref{fig:syn_vel_model}. $V_p$ was fixed while the corresponding $V_s$ values were calculated using the empirical relationship $V_p/V_s = 1.7$~\citep{brocher2005empirical}. $\rho$ was estimated using $\rho = 310 V_p^{1/4}$~\citep{gardner1974formation}. The model size was set based on the field data, with 3 m in both the horizontal and vertical directions. 

\begin{figure}
    \centering
    \includegraphics[width=1\linewidth]{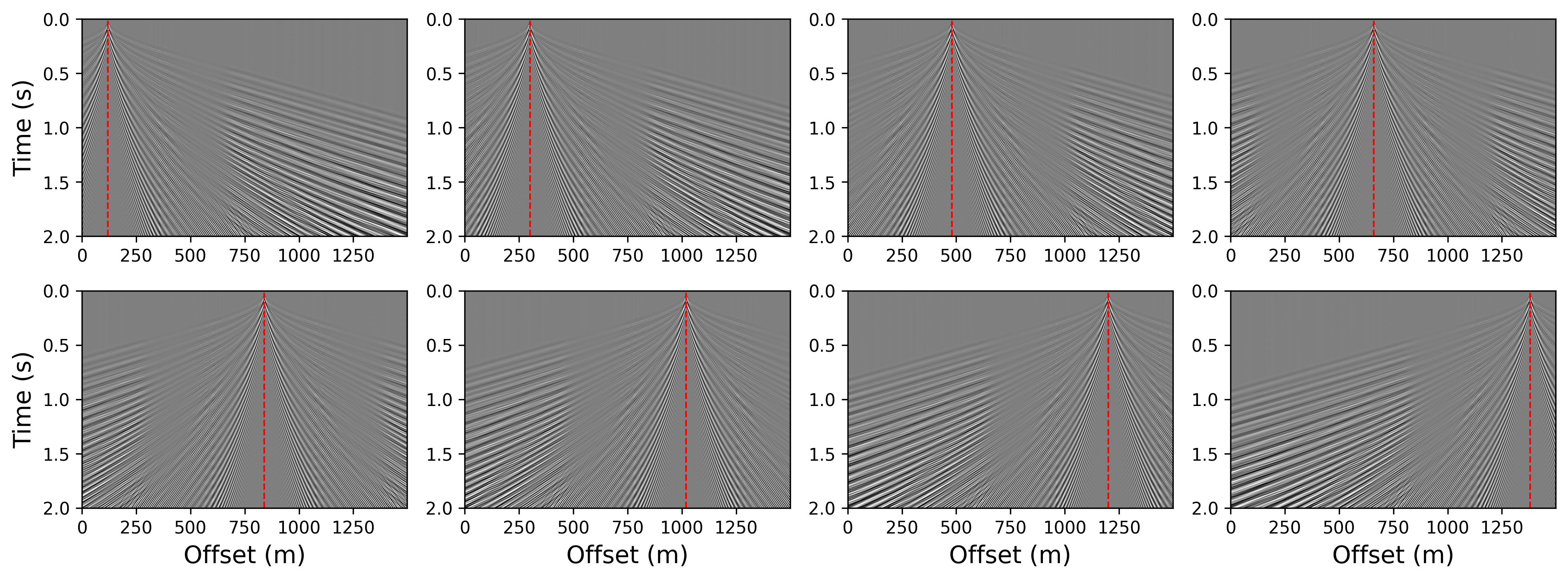}
    \caption{Simulated vertical-component shot gathers for eight surface sources with 25 Hz central frequency. The red dashed line indicates the source position for each shot.}
    \label{fig:shot_gathers}
\end{figure}

 Regarding the source, we used 8 source points evenly distributed on the surface. Two different tests were performed using a Ricker wavelet with central frequencies of 18 Hz and 25 Hz, while a time step of 0.1 ms and a total recording time of 2 s were used in all cases. We also used free-surface conditions on the top of the velocity model, and 120 absorbing layers were applied to the other boundaries. Figure~\ref{fig:shot_gathers} shows the resulting simulated shot gathers.

\begin{figure}
    \centering
    \includegraphics[width=1\linewidth]{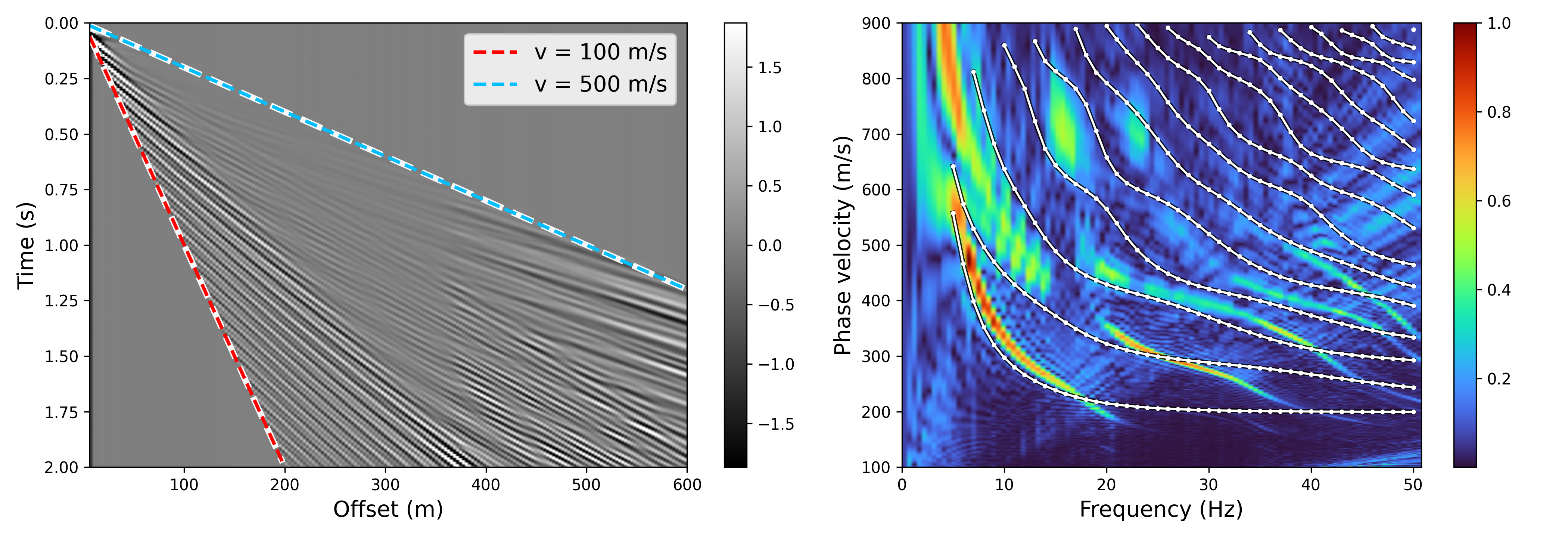}
    \caption{Synthetic vertical-component gather and its dispersion validation. The left panel shows the shot gather with a 100--500 m/s velocity muting, and the right panel compares the normalized $f$--$v$ dispersion image with theoretical multimode Rayleigh-wave curves.}
    \label{fig:syn_compare}
\end{figure}

To verify that the synthetic data contain physically meaningful multimode SW dispersion energy before network training and evaluation, Figure~\ref{fig:syn_compare} compares the simulated shot gather, its dispersion image, and theoretical Rayleigh-wave DCs, estimated for the 1D layered model of Figure~\ref{fig:syn_vel_model} using the  Haskell–Thomson transfer matrix method~\citep{thomson1950transmission,haskell1964radiation, haskell1990dispersion}. Figure~\ref{fig:syn_compare}a shows the vertical-component shot gather used for this comparison, after a 100--500 m/s velocity fan window (dashed lines in Figure~\ref{fig:syn_compare}, left). Velocity muting is applied to avoid simulation artifacts and to encourage the network to focus on the SW energy. The right panel in Figure~\ref{fig:syn_compare} shows the corresponding normalized dispersion image, where energy maxima are in good agreement with the fundamental-mode branch of the theoretical DC (white lines in Figure~\ref{fig:syn_compare}, right). In addition, weaker high-velocity energy branches appear in the dispersion image, corresponding to higher modes. This agreement confirms the physical consistency of the synthetic simulation.

\begin{figure}
    \centering
    \includegraphics[width=1\linewidth]{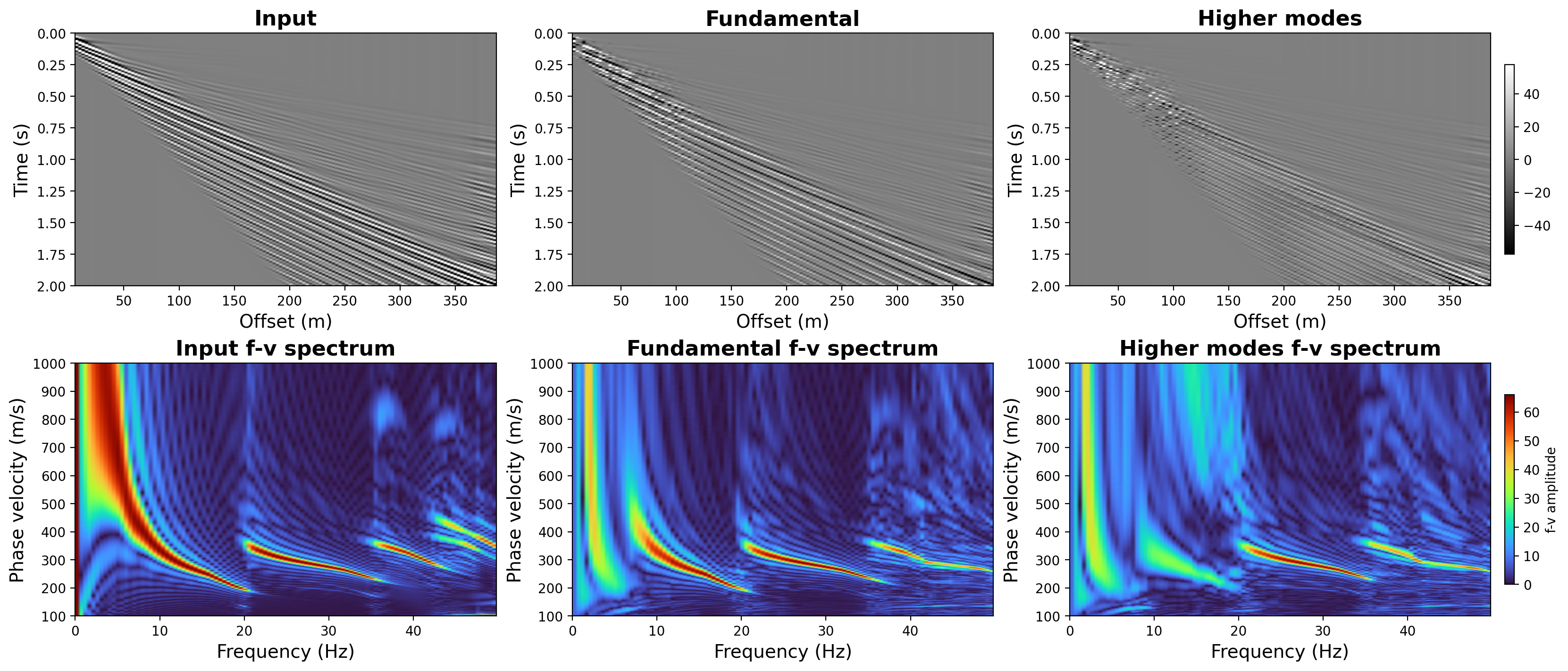}
    \caption{Mode separation results on synthetic data with a dominant central frequency of 25 Hz. First row (from left to right): input shot gather, output fundamental mode component, output higher-mode component. Second row (from left to right): corresponding dispersion images of the first row. Note that we applied velocity muting with lower and higher velocities of 100 m/s and 500 m/s, respectively, to avoid numerical artifacts.}
    \label{fig:mode_sep_syn_25hz}
\end{figure}

 Figure~\ref{fig:mode_sep_syn_25hz} displays the mode-separation results for the synthetic shot gather with a dominant central frequency of 25 Hz. After separation, the predicted fundamental-mode component mainly preserves coherent low-velocity, high-amplitude arrivals, whereas the residual output contains the remaining events. The corresponding $f$--$v$ dispersion images confirm the results: the fundamental-mode output concentrates energy along the dominant fundamental dispersion ridge, while the residual spectrum retains higher-mode energy at larger phase velocities and higher frequencies. These results indicate that the proposed physics-aware network can separate the fundamental mode from the higher modes while maintaining consistency with the input shot gather.

\begin{figure}
    \centering
    \includegraphics[width=1\linewidth]{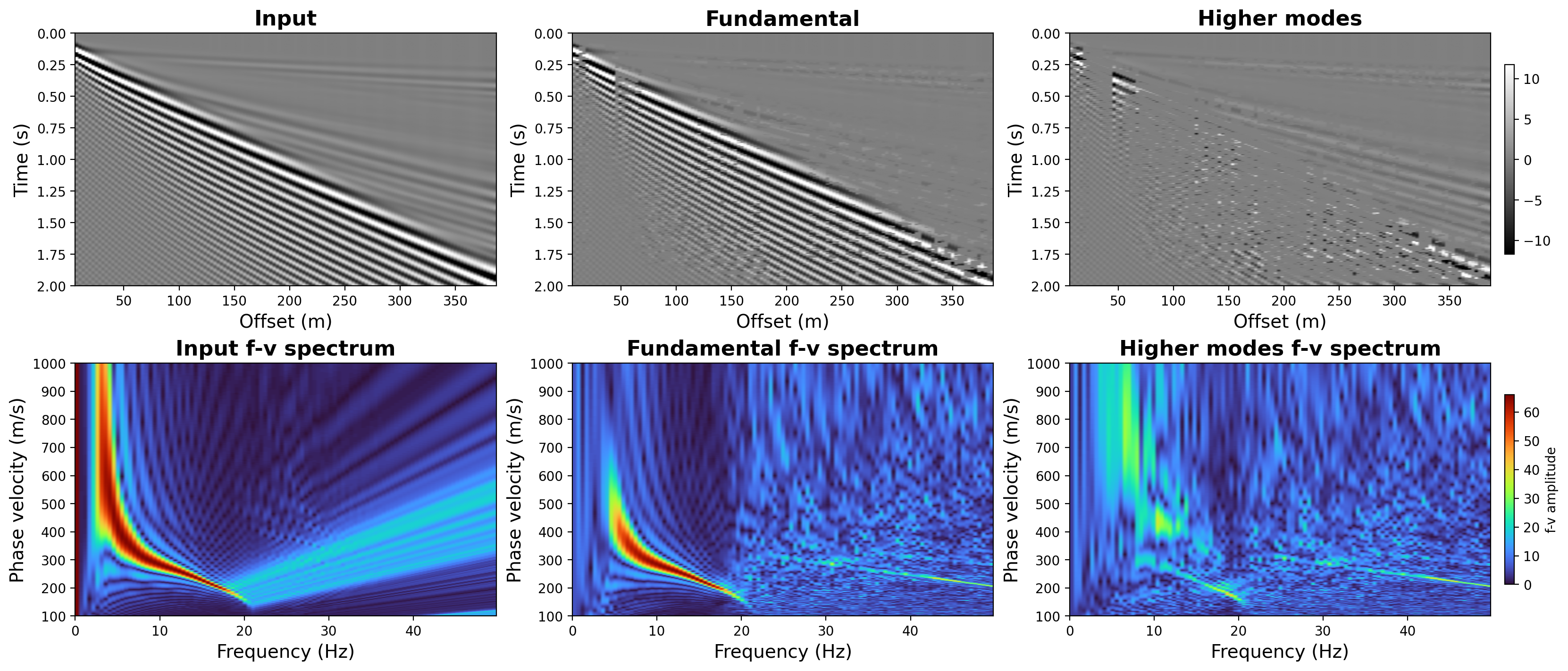}
    \caption{Mode separation results on synthetic data with a dominant central frequency of 18 Hz. First row (from left to right): input shot gather, output fundamental mode component, output higher-mode component. Second row (from left to right): corresponding dispersion images of the first row.}
    \label{fig:mode_sep_syn_18Hz}
\end{figure}

Figure~\ref{fig:mode_sep_syn_18Hz} shows the performance of mode separation on synthetic data with a central frequency of 18 Hz. Interestingly, as shown in Figure~\ref{fig:mode_sep_syn_18Hz}, the $f$--$v$ dispersion images reveal that the higher-mode energy becomes more distinguishable after separation. In the original data (left panel in Figure ~\ref{fig:mode_sep_syn_18Hz}), the higher-mode energy is largely masked due to the energetic dominance of the fundamental mode. After separation, the fundamental mode mainly preserves the strong low-velocity dispersion ridge, whereas the residual output enhances the weaker higher-mode energy that is difficult to identify in the input dispersion image. This finding indicates that the network not only separates the fundamental mode from the input but also recovers the physically meaningful higher-mode information.

\subsection{Field applications}
\begin{figure}
    \centering
    \includegraphics[width=0.85\linewidth]{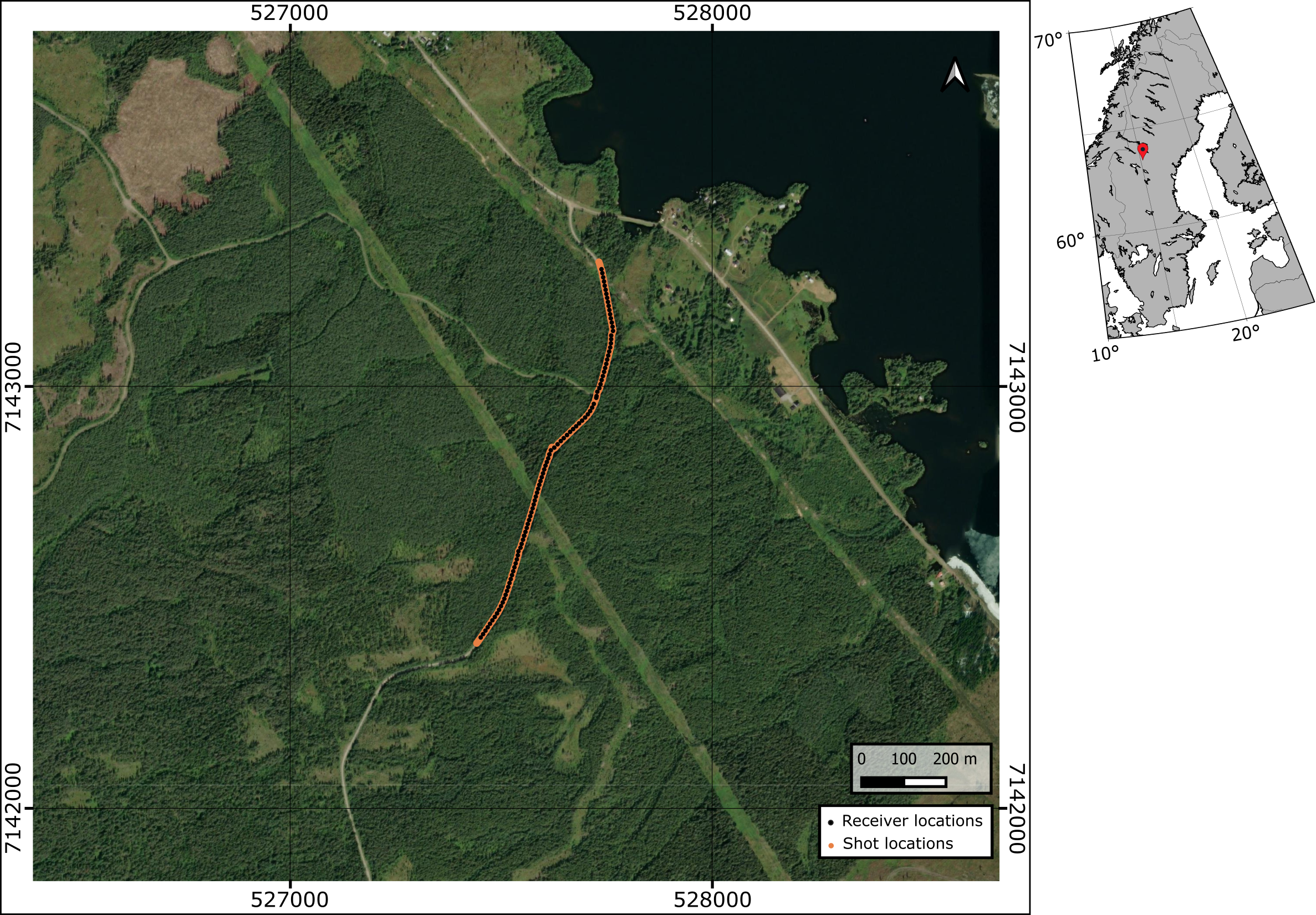}
    \caption{Geological map of the Alum Shale seismic survey acquired in August 2025. The main panel shows the survey area located in central Sweden, with receiver locations marked by black dots and sources as orange.}
    \label{fig:geo_map}
\end{figure}

The field data in Figure \ref{fig:geo_map} were recorded in August 2025 as part of a field survey in the Alum Shale Formation in central Sweden, near the village of Norråker. The Alum Shale Formation extends across much of Scandinavia, from northern Norway to southern Sweden~\citep{andersson1985scandinavian,gee2020lower}. It is locally enriched in uranium (U), vanadium (V), molybdenum (Mo), nickel (Ni), copper (Cu), and rare earth elements, making it a long-standing target for resource exploration~\citep{andersson1985scandinavian,vine1970geochemistry}. In addition to its polymetallic potential, the Alum Shale Formation has historically been exploited as an oil shale because of its significant hydrocarbon-generating capacity~\citep{andersson1985scandinavian,sanei2014petrographic}.

\begin{figure}
    \centering
    \includegraphics[width=1\linewidth]{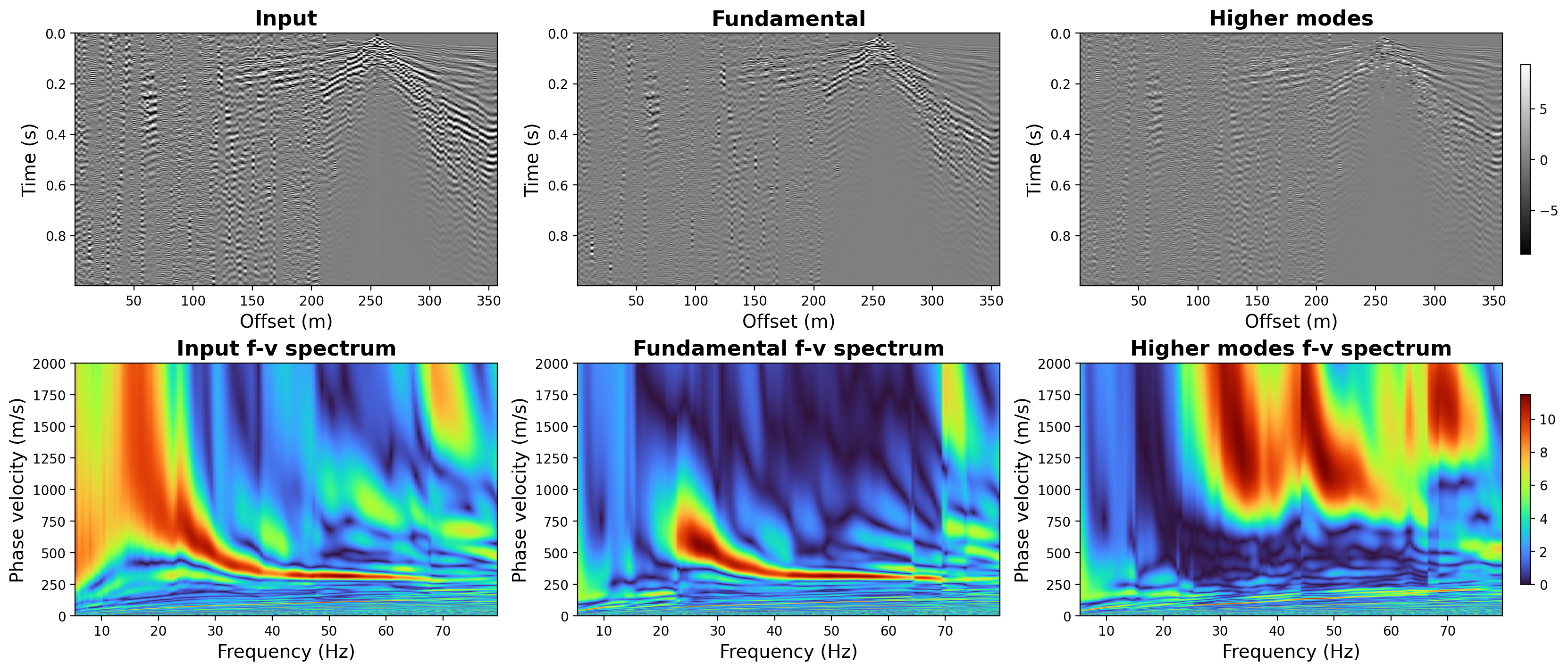}
    \caption{Mode separation results for field example. First row (from left to right): input shot gather, output fundamental mode component, output higher-mode component. Second row (from left to right): corresponding $f$--$v$ dispersion images of the first row.}
    \label{fig:station_85_mode_sep}
\end{figure}

Figure~\ref{fig:geo_map} displays the profile for our survey, with a NE--SW extension, which contains 98 3-component nodal receivers (marked by black dots), deployed in two sets of 49 receivers with~10 m receiver spacing, and 317 shot points with~3.33 m spacing (orange dots), with 157 shots for the first deployment set and 160 shots for the second deployment. A hammer was used as the source, excited on bedrock or boulders along the profile. For each shot location, we generated three shots which were stacked during processing to enhance the signal-to-noise ratio. The data were recorded at a 1 ms sampling rate in continuous mode and later reduced to 2.4 s. 
For SW processing, the data were transformed in the station-gather domain (98 station gathers in total), which, consistent with the findings of ~\cite{gomo2026near}, improved spatial resolution given the acquisition design and layout. As seen in Figure~\ref{fig:station_85_mode_sep}, compared to the synthetic examples, the field station gathers contain stronger noise and irregular amplitude variations, making them more challenging for mode separation. 

Figure~\ref{fig:station_85_mode_sep} shows the mode-separation results for one of the station gathers. In the input data (Figure~\ref{fig:station_85_mode_sep}, left), SW energy is strongly mixed, and different modal components overlap in both the space-time and $f$--$v$ domains. After separation, the predicted fundamental-mode component preserves the dominant coherent dispersive events, which are primarily associated with the low-velocity energy ridge in the $f$--$v$ dispersion images. In contrast, the higher-mode output contains the remaining modal energy, including stronger high-velocity components that are less evident in the input spectrum. The corresponding $f$--$v$ dispersion images demonstrate that the proposed method can redistribute the mixed SW energy into physically interpretable fundamental and higher-mode components, even for field data with complex wavefield characteristics. 

\begin{figure}
    \centering
    \includegraphics[width=1\linewidth]{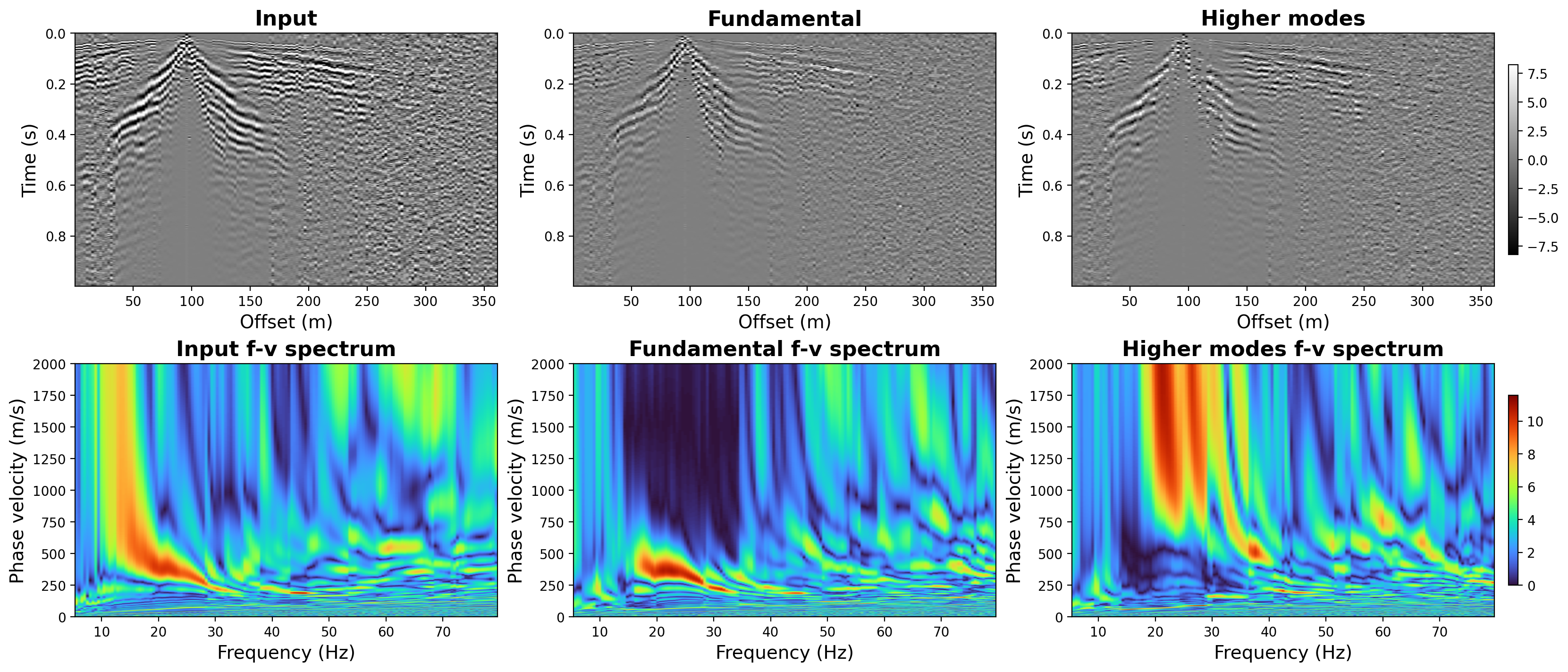}
    \caption{Mode separation results for field example. First row (from left to right): input shot gather, output fundamental mode component, output higher-mode component. Second row (from left to right): corresponding $f$--$v$ dispersion images of the first row.}
    \label{fig:station_13_mode_sep}
\end{figure}

An additional example in Figure~\ref{fig:station_13_mode_sep}, with more irregular amplitude variations and a more complex multimode pattern in the $f$--$v$ domain, further validates the proposed method's performance. 

\section{Discussion}
The proposed framework achieves promising SW mode separation results on both synthetic and field data, demonstrating the effectiveness of using DL as an inverse transform operator. Improving mode separation is important for two-station SW analysis, as cleaner input from two stations leads to more reliable DC picking. To further investigate the behavior, robustness, and reliability of the network, this section presents experiments on the influence of two-station DCs and uncertainty quantification. Based on the current findings, we also provide practical recommendations for applying the proposed framework to SW mode separation.

\subsection{Impact on 2-station analysis}
A major motivation for SW mode separation is to improve the quality of input data for two-station SW analysis. Unlike MASW, the two-station method estimates direct path-averaged dispersion between station pairs and is therefore allowing the tomography scheme to be applied for high resolution~\citep{kastle2016two}. To demonstrate this benefit, we compare the two-station dispersion measurements obtained from the original data and the separated components. 

\begin{figure}
    \centering
    \includegraphics[width=0.8\linewidth]{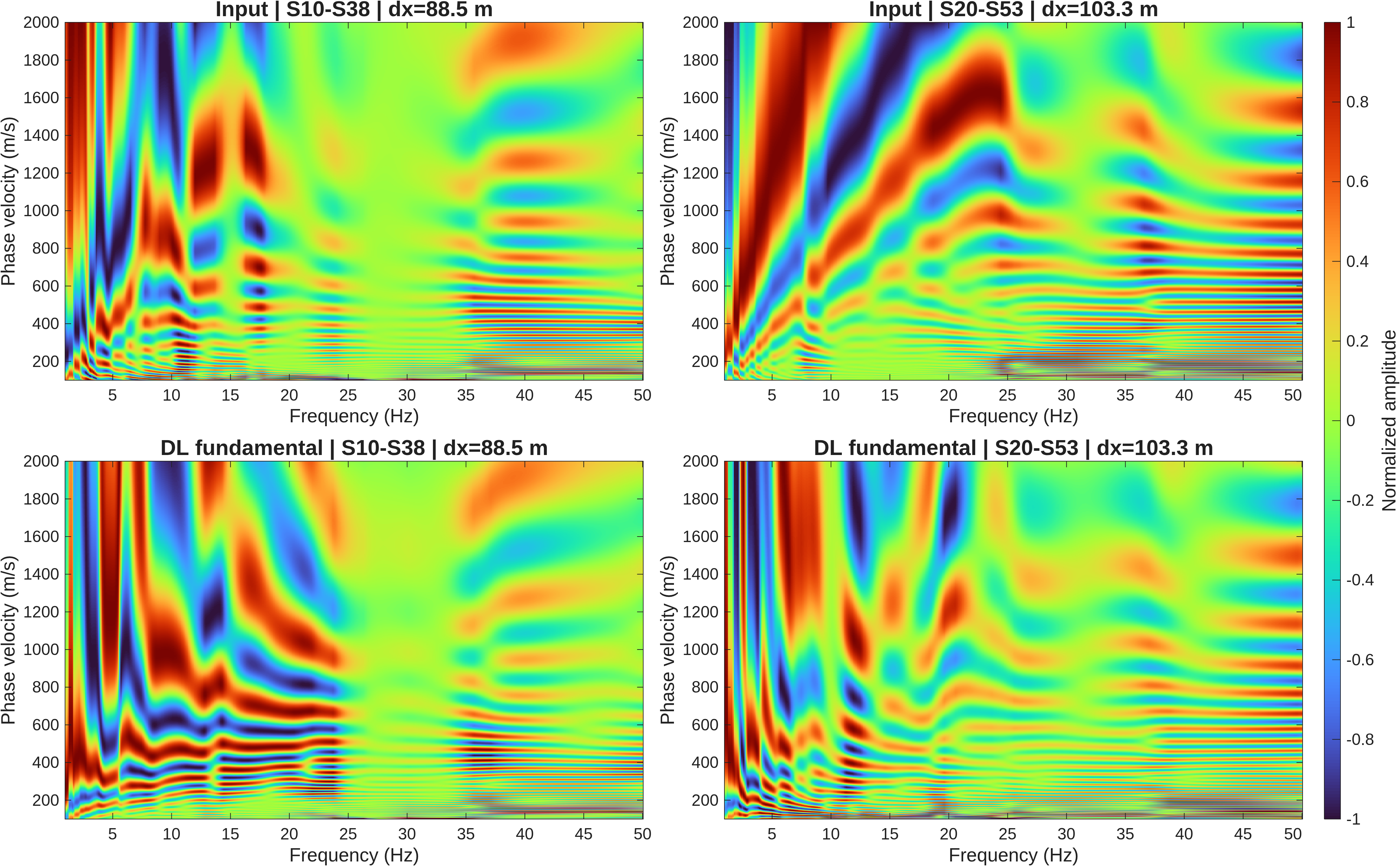}
    \caption{Two-station dispersion spectra computed from the original data shown in Figure~\ref{fig:station_85_mode_sep} and from the deep-learning-separated fundamental-mode component. The comparisons are shown for two two-station pairs along the profile: stations 10-38 and stations 20-58.}
    \label{fig:2-station_comparison}
\end{figure}

Figure~\ref{fig:2-station_comparison} compares the two-station spectra obtained from the original mixed-mode field data and the DL-separated data. Two two-station pairs, S10--S38 and S20--S58, are randomly selected for comparison. For both pairs, the spectra computed from the mixed-mode data exhibit stronger modal interference and less continuous dispersion ridges. After separation, the input data mainly contain the fundamental-mode energy, and the resulting spectra become more coherent and better concentrated, particularly in the lower-velocity range. These improvements facilitate more reliable two-station DC picking than the mixed-mode input does, thereby providing greater confidence in subsequent two-station dispersion analysis.

\subsection{Uncertainty quantification}
To mitigate the effects of non-uniqueness and assess the reliability of the mode-separation results, we employ mask-perturbation uncertainty as an uncertainty quantification (UQ) strategy~\citep{lakshminarayanan2017simple,cui2025learning}. UQ is important for downstream two-station dispersion analysis and SW inversion because uncertainty in separation results can propagate into biased dispersion measurements and inaccurate $V_s$ models.

We first conduct the mask-perturbation UQ analysis on the 25 Hz synthetic data, using the same setup as in Figure~\ref{fig:mode_sep_syn_25hz}. The top row compares the input gather, the mean DL fundamental-mode result, and the standard deviation caused by perturbing the dispersion mask. The uncertainty is mainly concentrated along the energetic SW arrivals, indicating that the separated waveform is most sensitive to mask changes where different modes overlap. The $f$--$v$ dispersion images standard deviation highlights that the separation depends strongly on the assumed mask, especially near overlapping modal energy and higher-frequency branches. 
\begin{figure}
    \centering
    \includegraphics[width=1\linewidth]{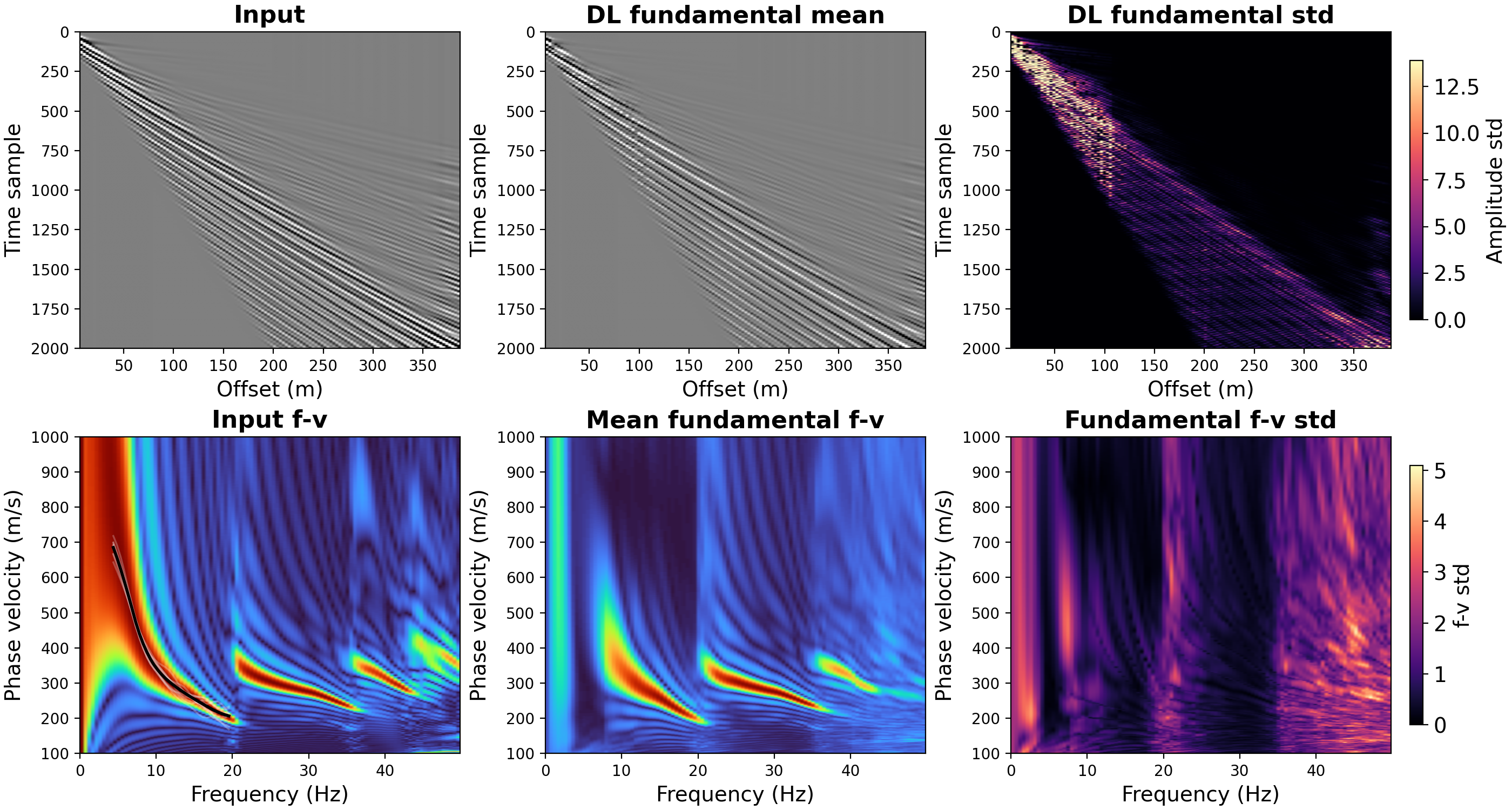}
    \caption{Uncertainty in the DL-separated fundamental mode estimated by perturbing the dispersion mask. The standard deviation shows regions where the separated result is sensitive to the assumed dispersion constraint.}
    \label{fig:uq_mask}
\end{figure}

\subsection{Limitations and future work}
Overall, the proposed method achieves promising results on both synthetic and field data without requiring labels for training. Nevertheless, several limitations remain in the current study. For synthetic data, we apply automatic dispersion-curve picking based on the energy maxima in the $f$--$v$ domain. However, this strategy may fail when different mode components overlap within the same frequency range. Therefore, manually picked DCs are used for the field data examples. In addition, for field applications, several parameters still need to be fine-tuned according to the energy distribution of the fundamental mode, including the width of the Gaussian mask and the cutoff frequency band used in the physics-aware loss function.

Future work can be extended in several directions. First, adaptive strategies can be developed to automatically determine the loss weights or loss terms based on the mask or the estimated fundamental-mode dispersion information. Second, stable and robust automatic DC picking methods, potentially based on machine learning, should be explored for field data. Third, the proposed physics-aware framework can be further generalized to other geophysical or image-processing tasks, where different signal components need to be separated in transformed domains.

\section{Conclusions}
We proposed a dispersion-guided, physics-aware deep inverse operator for SW mode separation. The method separates fundamental- and higher-mode components directly from mixed-mode shot gathers without requiring labeled training data. By using an adaptive Gaussian mask in the $f$--$v$ domain, the proposed loss function promotes the separation of the wavefield to preserve the target modal energy while suppressing energy leakage outside the prescribed dispersion region. Synthetic and field examples show that the proposed framework can reduce modal interference and produce cleaner fundamental-mode shot gathers. Mode separation results further emphasize that the framework is applicable to both synthetic and field data, thereby improving two-station or MASW $f$--$v$ dispersion images. Future work will focus on applying the separated results to subsequent two-station dispersion analysis and surface-wave tomography.

\section{Code and Resources}
The code associated with this study, including the implementation of the proposed method and the SW simulation codes, will be made publicly available at \url{https://github.com/cuiyang512/Deep-Inverse-Operator} upon acceptance of the paper.

\section{ACKNOWLEDGMENTS}
This work is supported by the Smart Exploration Research Center. The center has received funding from the Swedish Foundation for Strategic Research (SSF) under grant agreement no. CMM22-0003. This is publication SE26-049. The authors would like to thank Dr. Denis Anikiev (KFUPM), Dr. Umair Bin Waheed (KFUPM), and Dr. Ayse Kaslilar (Uppsala University) for their constructive discussions and valuable suggestions on the design of the physics-aware loss function and the synthetic data simulation. They also thank Mr. Csaba Gyuris (Eötvös Loránd University and Uppsala University) for providing the manually picked DCs for field data, and Dr. Iwona Klonowska (Uppsala University) for financing, co-organizing, and participating in the Alum Shale field campaign. 

\bibliographystyle{unsrtnat}
\bibliography{references}

@inproceedings{ronneberger2015u,
  title={U-net: Convolutional networks for biomedical image segmentation},
  author={Ronneberger, Olaf and Fischer, Philipp and Brox, Thomas},
  booktitle={International Conference on Medical Image Computing and Computer-Assisted Intervention},
  pages={234--241},
  year={2015},
  organization={Springer}
}

@Article{chao20252,
  author={Chao Li and Omar M. Saad and Yangkang Chen},
  title = {Unsupervised deep learning for off-the-grid seismic reconstruction and denoising},
    journal={Geophysics},
  year=2025,
  volume=64,
  issue=1,
  number=1,
  pages={doi: 10.1190/geo2024-0189.1},
  doi={10.1190/geo2024-0189.1},
}

@Article{geofwi,
  author={Chao Li and Yiran Shen and Sergey Fomel and Umair bin Waheed and Alexandros Savvaidis and Yangkang Chen},
  title = {GeoFWI: A large velocity model dataset for benchmarking full waveform inversion using deep learning},
    journal={Journal of Geophysical Research: Machine Learning and Computation},
  year=2026,
  volume=3,
  issue=1,
  number=1,
  pages={e2025JH001037},
  doi={10.1029/2025JH001037},
}

@Article{shucai2020,
  author={Shucai Li and Bin Liu and Yuxiao Ren and Yangkang Chen and Senlin Yang and Yunhai Wang and Peng Jiang},
  title = {Deep learning Inversion of Seismic Data},
  journal={IEEE Transactions on Geoscience and Remote Sensing},
  year=2020,
  volume=58,
  number=3,
  issue=3,
  pages={2135-2149},
  doi={10.1109/TGRS.2019.2953473},
}

@Article{omar2020ddae,
  author={Omar M. Saad and Yangkang Chen},
  title = {Deep Denoising Autoencoder for Seismic Random Noise Attenuation},
  journal={Geophysics},
  year=2020,
  volume=85,
  issue=4,
  number=4,
  pages={V367–V376},
  doi={10.1190/geo2019-0468.1},
}

@article{cui2025learning,
  title        = {Learning from Imperfect Labels: A Physics-Aware Neural Operator with Application to DAS Data Denoising},
  author       = {Cui, Yang and Anikiev, Denis and Bin Waheed, Umair and Chen, Yangkang},
  journal      = {arXiv preprint arXiv:2511.15638},
  year         = {2025},
  url          = {https://arxiv.org/abs/2511.15638}
}

@article{dorman1962numerical,
  title={Numerical inversion of seismic surface wave dispersion data and crust-mantle structure in the New York-Pennsylvania area},
  author={Dorman, James and Ewing, Maurice},
  journal={Journal of Geophysical Research},
  volume={67},
  number={13},
  pages={5227--5241},
  year={1962},
  publisher={Wiley Online Library}
}

@incollection{park1998imaging,
  title={Imaging dispersion curves of surface waves on multi-channel record},
  author={Park, Choon Byong and Miller, Richard D and Xia, Jianghai},
  booktitle={SEG technical program expanded abstracts 1998},
  pages={1377--1380},
  year={1998},
  publisher={Society of Exploration Geophysicists}
}

@article{xia1999estimation,
  title={Estimation of near-surface shear-wave velocity by inversion of Rayleigh waves},
  author={Xia, Jianghai and Miller, Richard D and Park, Choon B},
  journal={geophysics},
  volume={64},
  number={3},
  pages={691--700},
  year={1999},
  publisher={Society of Exploration Geophysicists}
}

@article{park1999multichannel,
  title={Multichannel analysis of surface waves},
  author={Park, Choon B and Miller, Richard D and Xia, Jianghai},
  journal={Geophysics},
  volume={64},
  number={3},
  pages={800--808},
  year={1999},
  publisher={Society of Exploration Geophysicists}
}

@article{socco2010surface,
  title={Surface-wave analysis for building near-surface velocity models—Established approaches and new perspectives},
  author={Socco, Laura Valentina and Foti, Sebastiano and Boiero, Daniele},
  journal={Geophysics},
  volume={75},
  number={5},
  pages={75A83--75A102},
  year={2010},
  publisher={Society of Exploration Geophysicists}
}

@article{xia2003inversion,
  title={Inversion of high frequency surface waves with fundamental and higher modes},
  author={Xia, Jianghai and Miller, Richard D and Park, Choon B and Tian, Gang},
  journal={Journal of Applied Geophysics},
  volume={52},
  number={1},
  pages={45--57},
  year={2003},
  publisher={Elsevier}
}

@article{pan2019sensitivity,
  title={Sensitivity analysis of dispersion curves of Rayleigh waves with fundamental and higher modes},
  author={Pan, Lei and Chen, Xiaofei and Wang, Jiannan and Yang, Zhentao and Zhang, Dazhou},
  journal={Geophysical Journal International},
  volume={216},
  number={2},
  pages={1276--1303},
  year={2019},
  publisher={Oxford University Press}
}

@article{luo2008rayleigh,
  title={Rayleigh-wave dispersive energy imaging and mode separating by high-resolution linear Radon transform},
  author={Luo, Yinhe and Xu, Yixian and Liu, Qingsheng and Xia, Jianghai},
  journal={The Leading Edge},
  volume={27},
  number={11},
  pages={1536--1542},
  year={2008},
  publisher={Society of Exploration Geophysicists}
}

@article{li2020separation,
  title={Separation of multi-mode surface waves by supervised machine learning methods},
  author={Li, Jing and Chen, Yuqing and Schuster, Gerard T},
  journal={Geophysical Prospecting},
  volume={68},
  number={4},
  pages={1270--1280},
  year={2020},
  publisher={European Association of Geoscientists \& Engineers}
}

@article{yao2006surface,
  title={Surface-wave array tomography in SE Tibet from ambient seismic noise and two-station analysis—I. Phase velocity maps},
  author={Yao, Huajian and van Der Hilst, Robert D and De Hoop, Maarten V},
  journal={Geophysical Journal International},
  volume={166},
  number={2},
  pages={732--744},
  year={2006},
  publisher={Blackwell Publishing Ltd Oxford, UK}
}

@article{yao2008surface,
  title={Surface wave array tomography in SE Tibet from ambient seismic noise and two-station analysis-II. Crustal and upper-mantle structure},
  author={Yao, Huajian and Beghein, Caroline and Van Der Hilst, Robert D},
  journal={Geophysical Journal International},
  volume={173},
  number={1},
  pages={205--219},
  year={2008},
  publisher={Blackwell Publishing Ltd Oxford, UK}
}

@article{gosselin2023probabilistic,
  title={Probabilistic inversion of circular phase spectra: application to two-station phase-velocity dispersion estimation in western Canada},
  author={Gosselin, Jeremy M and Audet, Pascal and Est{\`e}ve, Cl{\'e}ment and Schaeffer, Andrew J},
  journal={Geophysical Journal International},
  volume={233},
  number={2},
  pages={1387--1398},
  year={2023},
  publisher={Oxford University Press}
}

@article{schonewille2001parabolic,
  title={Parabolic Radon transform, sampling and efficiency},
  author={Schonewille, MA and Duijndam, AJW},
  journal={Geophysics},
  volume={66},
  number={2},
  pages={667--678},
  year={2001},
  publisher={Society of Exploration Geophysicists}
}

@phdthesis{papadopoulou2021surface,
  author = {Papadopoulou, Myrto},
  title  = {Surface-wave methods for mineral exploration},
  school = {Politecnico di Torino},
  year   = {2021},
  type   = {Doctoral thesis}
}

@article{jin2025extracting,
  title={Extracting Multimodal Surface-Wave Dispersion Curves from Ambient Seismic Noise Using High-Resolution Linear Radon Transform},
  author={Jin, Hao and Luo, Yinhe and Yang, Yingjie and Zhao, Kaifeng},
  journal={Seismological Research Letters},
  volume={96},
  number={1},
  pages={270--281},
  year={2025},
  publisher={Seismological Society of America}
}

@article{mousavi2022deep,
  title={Deep-learning seismology},
  author={Mousavi, S Mostafa and Beroza, Gregory C},
  journal={Science},
  volume={377},
  number={6607},
  pages={eabm4470},
  year={2022},
  publisher={American Association for the Advancement of Science}
}

@article{chamorro2024deep,
  title={Deep learning-based extraction of surface wave dispersion curves from seismic shot gathers},
  author={Chamorro, Danilo and Zhao, Jiahua and Birnie, Claire and Staring, Myrna and Fliedner, Moritz and Ravasi, Matteo},
  journal={Near Surface Geophysics},
  volume={22},
  number={4},
  pages={421--437},
  year={2024},
  publisher={European Association of Geoscientists \& Engineers}
}

@article{haskell1990dispersion,
  title={The dispersion of surface waves on multilayered media},
  author={Haskell, Norman A},
  journal={Vincit veritas: A portrait of the life and work of Norman Abraham Haskell, 1905--1970},
  volume={30},
  pages={86--103},
  year={1990},
  publisher={Wiley Online Library}
}

@article{gomo2026near,
  title={Near-surface characterization and delineation of water preferential flow-pathways at South Deep Gold Mine, South Africa},
  author={Gomo, Sikelela and Khosro Anjom, Farbod and Colombero, Chiara and Karimpour, Mohammadkarim and Jogee, Bibi Ayesha and Manzi, Musa SD and Socco, Laura V},
  journal={Solid Earth},
  volume={17},
  number={2},
  pages={249--273},
  year={2026},
  publisher={Copernicus GmbH}
}

@article{kastle2016two,
  title={Two-receiver measurements of phase velocity: cross-validation of ambient-noise and earthquake-based observations},
  author={K{\"a}stle, Emanuel D and Soomro, Riaz and Weemstra, Cornelis and Boschi, Lapo and Meier, Thomas},
  journal={Geophysical Journal International},
  volume={207},
  number={3},
  pages={1493--1512},
  year={2016},
  publisher={Oxford University Press}
}

@article{lakshminarayanan2017simple,
  title={Simple and scalable predictive uncertainty estimation using deep ensembles},
  author={Lakshminarayanan, Balaji and Pritzel, Alexander and Blundell, Charles},
  journal={Advances in neural information processing systems},
  volume={30},
  year={2017}
}

@techreport{andersson1985scandinavian,
  title        = {The Scandinavian Alum Shales},
  author       = {Andersson, A. and Dahlman, B. and Gee, D. G. and Sn{\"a}ll, S.},
  year         = {1985},
  institution  = {Sveriges Geologiska Unders{\"o}kning},
  series       = {Serie Ca},
  number       = {56},
  pages        = {1--50},
  address      = {Uppsala}
}

@incollection{gee2020lower,
  title     = {Lower thrust sheets in the Caledonide orogen, Sweden: Cryogenian--Silurian sedimentary successions and underlying, imbricated, crystalline basement},
  author    = {Gee, David G. and Stephens, Michael B.},
  booktitle = {Sweden: Lithotectonic Framework, Tectonic Evolution and Mineral Resources},
  editor    = {Stephens, Michael B. and Bergman, Stefan},
  series    = {Geological Society, London, Memoirs},
  volume    = {50},
  chapter   = {20},
  pages     = {495--515},
  year      = {2020},
  publisher = {Geological Society of London},
  doi       = {10.1144/M50-2018-7}
}

@article{vine1970geochemistry,
  title={Geochemistry of black shale deposits; a summary report},
  author={Vine, James D and Tourtelot, Elizabeth B},
  journal={Economic Geology},
  volume={65},
  number={3},
  pages={253--272},
  year={1970},
  publisher={Society of Economic Geologists}
}

@article{sanei2014petrographic,
  title={Petrographic and geochemical composition of kerogen in the Furongian (U. Cambrian) Alum Shale, central Sweden: Reflections on the petroleum generation potential},
  author={Sanei, H and Petersen, HI and Schovsbo, NH and Jiang, C and Goodsite, Michael Evan},
  journal={International Journal of Coal Geology},
  volume={132},
  pages={158--169},
  year={2014},
  publisher={Elsevier}
}

@article{papadopoulou2023high,
  title={High-resolution static corrections derived from surface-wave tomography: Application to mineral exploration},
  author={Papadopoulou, Myrto and Brodic, Bojan and Koivisto, Emilia and Kaleshova, Albina and Savolainen, Mikko and Marsden, Paul and Socco, Laura Valentina},
  journal={Geophysics},
  volume={88},
  number={6},
  pages={B317--B328},
  year={2023},
  publisher={Society of Exploration Geophysicists}
}

@article{yang2026unsupervised,
  title={Unsupervised Physics-Guided Deconvolution for High-Resolution Hardrock Seismic Imaging},
  author={Yang, Liuqing and Malehmir, Alireza and Markovic, Magdalena},
  journal={Geophysical Prospecting},
  volume={74},
  number={1},
  pages={e70123},
  year={2026},
  publisher={European Association of Geoscientists \& Engineers}
}

@article{cercato2009addressing,
  title={Addressing non-uniqueness in linearized multichannel surface wave inversion},
  author={Cercato, Michele},
  journal={Geophysical Prospecting},
  volume={57},
  number={1},
  pages={27--47},
  year={2009},
  publisher={European Association of Geoscientists \& Engineers}
}

@book{foti2014surface,
  title={Surface wave methods for near-surface site characterization},
  author={Foti, Sebastiano and Lai, Carlo G and Rix, Glenn J and Strobbia, Claudio},
  year={2014},
  publisher={CRC press}
}

@article{ikeda2020two,
  title={Two-station continuous wavelet transform cross-coherence analysis for surface-wave tomography using active-source seismic data},
  author={Ikeda, Tatsunori and Tsuji, Takeshi},
  journal={Geophysics},
  volume={85},
  number={1},
  pages={EN17--EN28},
  year={2020},
  publisher={Society of Exploration Geophysicists}
}

@article{thomson1950transmission,
  title={Transmission of elastic waves through a stratified solid medium},
  author={Thomson, William T},
  journal={Journal of applied Physics},
  volume={21},
  number={2},
  pages={89--93},
  year={1950}
}

@article{haskell1964radiation,
  title={Radiation pattern of surface waves from point sources in a multi-layered medium},
  author={Haskell, NA},
  journal={Bulletin of the Seismological Society of America},
  volume={54},
  number={1},
  pages={377--393},
  year={1964},
  publisher={The Seismological Society of America}
}

@article{gardner1974formation,
  title={Formation velocity and density; the diagnostic basics for stratigraphic traps},
  author={Gardner, GHF and Gardner, LW and Gregory, ARw},
  journal={Geophysics},
  volume={39},
  number={6},
  pages={770--780},
  year={1974},
  publisher={Society of Exploration Geophysicists}
}

@article{brocher2005empirical,
  title={Empirical relations between elastic wavespeeds and density in the Earth's crust},
  author={Brocher, Thomas M},
  journal={Bulletin of the seismological Society of America},
  volume={95},
  number={6},
  pages={2081--2092},
  year={2005},
  publisher={Seismological Society of America}
}

\end{document}